\documentstyle[floats,prd,aps,epsf,12pt]{revtex}

\makeatletter
\newbox\tempboxa
\newdimen\captionboxsubcount
\def\capsize#1{\captionboxsubcount=#1pt}
\newdimen\captionboxsub
\captionboxsub=\hsize \advance\captionboxsub by -\captionboxsubcount
\advance\captionboxsub by -\captionboxsubcount
\long
\def\@makecaption#1#2{
\setbox\@tempboxa\hbox{#1 #2}
\ifdim \wd\@tempboxa >\captionboxsub
\rightskip=\captionboxsubcount \leftskip=\captionboxsubcount #1 #2
\else \hbox to\hsize{\hfil\box\@tempboxa\hfil}
\fi}
\makeatother
\capsize{30}

\begin{document}

\begin{titlepage}
\begin{flushright}
\begin{minipage}{5cm}
\begin{flushleft}
\small
\baselineskip = 13pt
YCTP-P?-99\\  hep-th/yymmnn \\
\end{flushleft}
\end{minipage}
\end{flushright}
\begin{center}
\Large\bf
Concatenation of Scales Below 1 eV \footnote{ Dedicated to Kurt
Haller on his seventieth birthday.}
\end{center}
\vfil

\begin{center}

Alan {\sc Chodos}\footnote{ Electronic address : {\tt
chodos@hepvms.physics.yale.edu}}\\ { \it
\qquad Center for Theoretical Physics}\\
{ \it Department of Physics, Yale University, New Haven, CT
06520-8120}\\

\qquad

\end{center}
\vfill
\begin{center}
\bf
Abstract
\end{center}
\begin{abstract} There are (at least) four numbers of physical and
cosmological significance, whose inferred values, when expressed in
mass units, cluster in a window below $1$ eV. There are: ~the
neutrino mass, the neutrino chemical potential, the cosmological
constant, and the size of two extra dimensions (if the fundamental
scale of gravity is $1-10$ TeV). In this note, we imagine ways in
which these four numbers could all be connected.

\baselineskip = 17pt

\end{abstract}
\begin{flushleft}

PACS numbers:
\end{flushleft}
\vfill
\end{titlepage}

\noindent

\bigskip
Numerology is not much in fashion these days. The grand tradition
of Kepler [1], Balmer [2], Eddington [3], and Dirac [4] has hardly
any followers, and even when one arises [5], he is quickly shot
down [6]. So it is with some trepidation that I want to bring up
the fact that at least four dimensionful numbers, all potentially
of great physical or cosmological significance, and all of which
have been discussed recently in the literature, seem to cluster in
the few decades just below one electron volt, which I shall loosely
refer to as "the window".

I shall endeavor to present a framework within which it can be
imagined that these four numbers might have something to do with
each other. Most of this is based on work that I did recently with
D. G. Caldi [7]; one component, in addition, is the subject of work
in progress with Erich Poppitz [8].

The four numbers in question are:

\bigskip\noindent
1) the neutrino mass. What one seems to deduce from a combination
of the solar and atmospheric neutrino data, with perhaps a dash of
LSND thrown in, is that neutrino mass differences lie inside this
window, although LSND prefers a rather higher value. The literature
on this topic is so extensive [9] that it would be redundant for me
to attempt to review it here.

\bigskip\noindent
2) the neutrino chemical potential [10]. Of course there are at
least 3 flavors of neutrino, and their chemical potentials may
differ. This quantity measures the difference between the
cosmological density of neutrinos and anti-neutrinos of whatever
species one is interested in. In fact, all of them may vanish, but
in any event, there is a sum rule that suggests that if they are
non-zero, at least some of them may lie inside the window of
interest.

\bigskip\noindent
3) the cosmological constant. Recently, interest in a non-zero
value of the cosmological constant has revived [11], spurred by
measurements of an accelerating universe. Of course the
acceleration may not be due to a strictly constant cosmological
constant as demanded by classical general relativity--there are
other ideas, such as "quintessence" [12] and the effects of
conformal gravity [13].  But what is true is that the required
energy density, expressed in mass units, falls squarely within the
window.

\bigskip\noindent
4) The magnitude of a pair of extra dimensions. Recently, inspired
by certain stringy ideas, it has been suggested [14] that the
fundamental scale for gravity is not the Planck scale, but rather
something in the range $1-10$ TeV (conveniently, and tantalizingly,
just above the reach of current accelerators). To achieve this, one
needs to imagine that whereas all the fields associated with the
standard model are confined to a 3-brane (i.e. four-dimensional
spacetime), gravity propagates in six dimensions, two of which are
compactified. The size of the compactified dimensions is determined
by the requirement that gravity is coupled to ordinary matter with
Newton's constant. This fixes the size to be about a millimeter
(for $1$ TeV) or a hundredth of that (for $10$ TeV), placing the
inverse size somewhere within the window.

What Caldi and I tried to do was to relate 1)-3) by hypothesizing a
neutrino condensate in the universe. We had in mind not a chiral
condensate of the type that would be familiar in the context of
QCD, but rather a condensate of Cooper pairs. The advantages of
this are twofold: first, the magnitude of such a condensate is tied
directly to the size of the Fermi surface, which in turn is
governed by the chemical potential; so 1) and 2) are immediately
related. Second, unlike the case of a chiral condensate, as long as
the chemical potential is non-zero, a pairing condensate will form
in any channel with an attractive interaction, no matter how small.

Our first thought was to generate the condensate within the
framework of the known neutrino interactions described by the
standard model. One integrates out the $W$ and $Z$, and expresses
this interaction in effective four-fermi form. Then one employs a
mean field approximation to generate a set of gap equations, and
one studies these to determine which are the attractive channels.

The simplest case is if one considers only one flavor of neutrino,
coupled to itself. Since one is using the standard model, the
neutrino is considered massless. The condensate itself, if any, is
what will generate the neutrino mass.

In this simple case, the gap equation takes the form

\begin{equation}
B = {- G^2 \over \pi^2} \int_{- \Lambda}^{\Lambda} ~dp ~p^2 ~{B
\over \sqrt{(p - \mu)^2 + 8 MG^2}} ~.
\end{equation}

Here $B$ is the value of $\langle \psi\psi \rangle$, (with spinor
indices suitably contracted), where $\langle ~\rangle$ denotes a
vacuum expectation value, and $M = B^{\dagger} B$. $G^2$  is the
effective four-fermi coupling, with dimension of inverse mass
squared, which in the standard model is related to the usual fermi
coupling constant $(10^{-5} M_{p}^{-2})$ by a coefficient of order
one. The parameter $\mu$ is the chemical potential, and $\Lambda$
is the cutoff, which is necessary since the effective four-fermi
theory is not renormalizable.

The important point to notice is that the two sides of this
equation are of opposite sign; this implies a repulsive channel,
and there is no solution to the gap equation other than the trivial
one $B = 0$.

It is not surprising that this channel is repulsive. After all, the
interaction is due to vector exchange, and as one learns from
electrodynamics, like particles repel. Since we have allowed only
one species of particle, it only interacts with itself, and hence
produces a repulsive interaction.

The next thing one might try is to introduce a variety of different
flavors. The condensate will then have a matrix structure in flavor
space:

\begin{equation}
B_{ij} = \langle \psi_i \psi_j \rangle ~,
\end{equation}

\bigskip\noindent
and one might hope that some of the elements of this matrix would effectively
represent attractive channels. But as long as the chemical potential remains
flavor independent, one can show with a little group theory   that the
problem essentially factorizes into a flavor part times a spinor part, and
the negative sign that was found for a single flavor persists. It is an
interesting question to see what might happen for the case of a flavor
dependent chemical potential, but Caldi and I did not examine that case,
although we did speculate that flavor dependent chemical potentials might
induce condensation in such a way as to generate an interesting spectrum
of neutrino masses and mixings.

As yet another, more radical possibility, one might consider the
pairing of neutrinos not with each other, but rather with charged
leptons. Phenomenologically, this possibility would be rather
drastic, because if such a condensate existed it would imply that
the vacuum was not an eigenstate of charge. In any case, one finds
that the pairing of a neutrino with a lepton of the same flavor
remains repulsive, but the pairing of a neutrino with a lepton of
different flavor (e.g. the muon neutrino with the electron) is
attractive in the standard model. Because the two members of the
pair have different mass, and because one must also allow different
chemical potentials for the neutrino and the charged lepton, one
obtains a rather more complicated form for the gap equation.
Explicitly, one finds

\begin{equation}
- \kappa^2 B^{\dagger}B = {- i \kappa^4 B^{\dagger}B \over \pi^3} \int_{-
\Lambda}^{\Lambda} ~dp ~p^2 \int_{- \infty}^{\infty} ~dp_0 {\cal G}(p_0, p)
\end{equation}

\bigskip\noindent
where

\begin{equation}
{\cal G}(p_0, p) = {p_0 - \mu_e + p \over [(p_0 - \mu_e)^2 - p^2 -
m^2][p_0 + \mu_{\nu}^{\prime} + p] - 4 \kappa^4 B^+ B [p_0 - \mu_e
+ p]} ~.
\end{equation}

Here $B = \langle \nu_{\mu}\gamma^0\gamma^2 e \rangle$ (a Lorentz
scalar), and $\kappa^2 = G^2 (1 - 2 sin^2 \theta_W)$, where
$\theta_W$ is the Weinberg angle. The parameters $\mu_e$ and
$\mu_{\nu}^{\prime}$ are the electron and neutrino chemical
potentials, respectively (the prime is to remind us that the
neutrino carries a different flavor quantum number from the charged
lepton).

It is not possible to simplify this expression much in the general
case. But it is instructive to look at some special limits, which I
do in the appendix. In particular, one learns that the property of
admitting a solution for arbitrarily weak attraction is maintained
only if the lepton and neutrino densities are equal and
non-vanishing, which is certainly not true of the present-day
universe.

Caldi and I also noted that there was the possibility of
lepton-neutrino condensation in a Lorentz-non-invariant channel,
but we did not investigate this.

The unrenormalized gap equations that I have introduced so far are
not suited to determining the overall size of the condensate,
because this will depend on the cutoff. Since the underlying theory
(the standard model) is renormalizable, presumably the cutoff
dependence can be consistently eliminated by introducing
renormalized parameters. One then expects that the expression for
the condensate will take a BCS-like form:

\begin{eqnarray}
\bigtriangleup = \kappa^2 B^{\dagger} B \sim \mu e^{- {1 \over \mu^2 G^2}} ~.
\end{eqnarray}

\bigskip\noindent
The condensate is proportional to the chemical potential, as
expected, but since the generation of a gap is a non-perturbative
phenomenon, one has the characteristic non-analytic dependence on
the coupling exhibited in eq. (2). The exponential reflects the
fact that as the gap tends to zero, the gap equation exhibits a
logarithmic singularity at the fermi surface. In our case, this
exponential factor is extremely small, pushing the condensate far
outside the desired window. This led Caldi and me to hypothesize
the existence of a new interaction, acting only among neutrinos,
which would effectively generate a four-fermi coupling with a scale
in the range between $1$ eV and $10^{-2}$ eV, instead of the $250$
GeV given by the standard model. Furthermore, of course, we assumed
the interaction to be attractive, effectively eliminating the sign
problem in eq. (1). Granting that this is ad hoc, let us just
pursue it a bit to see how the various quantities with which we
began enter the discussion.

Let us suppose that the dynamics of the neutrinos is governed by an
effective attractive four-fermi interaction of the form (in
two-component notation)

\begin{equation}
{\cal L}_{int} = - 2 G^2 (\psi_{\alpha}^{\dagger}
\epsilon_{\alpha\gamma} \psi_{\gamma}^{\dagger}) ~(\psi_{\beta}
\epsilon_{\beta\delta} \psi_{\delta})
\end{equation}

\bigskip\noindent
where the coupling G  is of the order an inverse eV. Furthermore, we
assume that there is a sufficiently large chemical potential to induce
condensation of neutrino pairs with a scale of an eV. This will
immediately give rise to a Majorana neutrino mass of magnitude

\begin{equation}
m = 2 G^2 \mid \langle \psi_{\alpha}^{\dagger}
\epsilon_{\alpha\gamma} \psi_{\gamma}^{\dagger} \rangle \mid ~.
\end{equation}

\bigskip\noindent
Furthermore, the condensate represents a contribution to the vacuum
energy density:

\begin{equation}
\epsilon_0 = - 2 G^2 \langle \psi_{\alpha}^{\dagger}
\epsilon_{\alpha\gamma} \psi_{\gamma}^{\dagger} \rangle \langle
\psi_{\beta} \epsilon_{\beta\delta}
\psi_{\delta} \rangle ~.
\end{equation}

\bigskip\noindent
According to the usual lore, however, this energy density will represent
an effective cosmological constant only if the system is not in the true
vacuum, but is rather in a supercooled state or perhaps in a very slowly
rolling state in which the size of the condensate is decaying toward
zero (the true vacuum). For definiteness, let us imagine the former
possibility. Then at some earlier time in the history of the universe,
the neutrinos condensed, and for a while the universe was in the true ground
state. As the universe evolved, the state with the condensate present
ceased to be the true ground state, so there can exist an inflationary
epoch (with a rather small cosmological constant) that persists until the
universe undergoes another phase transition to the true vacuum.

Does the above scenario make any sense when confronted with the
standard picture of how the universe is evolving? In the earlier
epoch, the background neutrinos were both at a higher density and a
higher temperature. As a consequence of its expansion, the universe
traces out a particular path as a function of time in the neutrino
temperature/density plane, heading for lower values of both
variables as time goes on. According to the standard big-bang
cosmology, $T \sim {1 \over R(t)}$ and $\rho \sim {1 \over
R^3(t)}$, so the curve in question is just $\rho \propto T^3$. Now
the Cooper pairing phase we have been discussing is a
low-temperature high-density phase. If the universe enters this
phase at all, one expects that it will eventually emerge from it as
the density drops. There is a program underway capable of testing
this expectation, in a one-plus-one dimensional simulation. This is
based on a recently introduced model [15] similar to the
Gross-Neveu model [16], which, however, also contains a Cooper-pair
type condensate as well as the usual chiral condensate. Once the
phase diagram of this model is understood, it will be possible to
trace out the history of the universe, to see whether there exists
an epoch such that the system is in a false vacuum characterized by
a neutrino condensate that will eventually disappear, along with
the attendant cosmological constant.

To summarize thus far: the formation of a cosmological condensate
of Cooper pairs would appear to require some new interaction acting
only among neutrinos. If this interaction is attractive and has the
right magnitude, and if the relevant chemical potentials are as
large as are allowed by the sum rules [10], and if we are at
present in a supercooled state, then both the neutrino mass and the
cosmological constant can be viewed as dynamical consequences of
this condensation phenomenon.

The reader may have been remarking at the high dosage of
speculation that has been necessary to get us to this point;
however, the most speculative part of this discussion is yet to
come. We shall adopt a recent idea [14]  motivated by certain
aspects of string theory: that the fundamental scale of quantum
gravity is not the Planck scale, but rather something not much
bigger than a TeV. If this is true, it opens up the exciting
possibility that quantum- gravitational effects will be accessible
to the next generation of accelerators. The price to be paid,
however, is that gravity does not propagate in four dimensions, but
rather in a six-dimensional spacetime, two of whose dimensions are
compactified into a manifold with the characteristic size of a
millimeter.

Now a millimeter is equivalent to about $10^{-4}$ eV, so this is
yet another scale that falls within our window. Is there any way of
tying it to the other three scales we have been discussing so far?
We begin by noting [8, 17] that there exists an exact classical solution of 
the six-dimensional Einstein equations with a cosmological constant
$\lambda_0$, which is the product of a four-dimensional de Sitter
space and a static 2-sphere. If the equation defining the de Sitter
space is

\begin{equation}
R_{\mu\nu} - {1 \over 2} g_{\mu\nu} R + \alpha g_{\mu\nu} = 0 ~,
\end{equation}

\bigskip\noindent
then the cosmological constant $\alpha$ is related to the bulk
cosmological constant by

\begin{equation}
\lambda_0 = 2 \alpha
\end{equation}

\bigskip\noindent
and to the radius of the $2$-sphere $R_0$ simply by

\begin{equation}
\alpha = {1 \over R_0^2} ~.
\end{equation}

\bigskip\noindent
However, the relationship between the energy density associated with the
observed cosmological constant and the postulated size of the extra
dimensions involves a power of the Planck scale, so this relationship is
violated by many orders of magnitude in nature (about 27).
Work is currently in progress [8] to find other
solutions of the six-dimensional equations, which include not only
gravity but also, at least partially, the effects of the matter
that resides on the four-dimensional ``$3$-brane" that we know as
spacetime, and which will involve so-called ``warp factors" as well.
It remains to be seen whether such modifications can lead to a realistic
geometry that incorporates both the observed cosmological constant and
a millimeter-sized pair of extra dimensions. 

The purpose of this article has been to weave some scenarios,
hopefully not too far outside the bounds of credibility, that might
begin to explain the remarkable concatenation of scales that one
observes in the mass region just below $1$ eV. The two ideas that
we have concentrated upon are, first, the existence of a
cosmological neutrino condensate of the Cooper pairing variety,
and, second, a cosmology governed by a geometry more complicated than but 
not too unlike the one described in eqns. (9-11).

\vfill\eject
\noindent
{\bf Appendix}

\bigskip\noindent
Here we examine special limits of the gap equation given by eqns.
(3) and (4). In particular, we shall briefly discuss the following:
(i) $m = 0$; (ii) $\mu_e = - \mu_{\nu}^{\prime} = \mu$; and (iii)
$B^{\dagger}B = 0$.

\bigskip\noindent
(i) When $m = 0$, we are considering the pairing of two massless
leptons. We therefore expect something similar to the gap equation
of eq. (1); the only difference is that we are still allowing them
to have unequal chemical potentials. We find that in this limit,
${\cal G}_0$ simplifies to:

\begin{equation}
{\cal G}_0(p_0, p) = {1 \over (p_0 - \mu_e - p)(p_0 +
\mu_{\nu}^{\prime} + p) - \Delta^2}
\end{equation}

\bigskip\noindent
where $\Delta^2 = 4 \kappa^2 B^{\dagger}B$. The $p_0$ integral can
then be done, giving for the gap equation

\begin{equation}
1 = {\kappa^2 \over \pi^2} \int_{- \Lambda}^{\Lambda} ~dp ~p^2
{\theta [(p + {\mu_e + \mu_{\nu}^{\prime} \over 2})^2
 +
\Delta^2
- ({\mu_e - \mu_{\nu}^{\prime} \over 2})^2] \over [(p + ({\mu_e +
\mu_{\nu}^{\prime} \over 2}))^2 + \Delta^2]^{{1 \over 2}}} ~.
\end{equation}

\bigskip\noindent
The $\theta$-function in the numerator arises from the $i\epsilon$
prescription $p_0 \rightarrow p_0 + i \epsilon sgn p_0$ that is
implicit in the expression for ${\cal G}_0$. We observe that,
first, this is an attractive channel since both sides of the gap
equation have the same sign. Second, without the $\theta$-function
(and modulo an inconsequential change in the overall sign of the
chemical potential) this is of the same general form as equation
(1). Third, the role of the $\theta$-function is to protect the
denominator from vanishing in the range of integration in the limit
$\Delta \rightarrow 0$. As $\mu_{\nu}^{\prime} - \mu_e \rightarrow
0$, this protection disappears, and one regains the instability at
the Fermi surface that one sees in equation (1).

\bigskip\noindent
(ii) When $\mu_{\nu}^{\prime} = - \mu_e = - \mu$, ${\cal G}_0$
becomes

\begin{equation}
{\cal G}_0(p_0, p) = {1 \over (p_0 - \mu)^2 - p^2 - m^2 - \Delta^2}
~.
\end{equation}

\bigskip\noindent
In this case, we see immediately that $\Delta^2$ acts like an extra
contribution to $m^2$, and doing the $p_0$ integral we obtain

\begin{equation} 1 = {\kappa^2 \over \pi^2}\int_{-
\Lambda}^{\Lambda} ~dp ~p^2 {\theta (p^2 + m^2 + \Delta^2 - \mu^2)
\over \sqrt{p^2 + m^2 +
\Delta^2}} ~.
\end{equation}

\bigskip\noindent
Here there is no possibility of an infrared instability, and one
will in general have a solution only if $\kappa$ is large enough.
This is exactly like a typical gap equation associated with a
chiral condensate, which is not surprising since by setting
$\mu_{\nu}^{\prime} = - \mu_e$ we are effectively treating one
member of the pair as a particle and the other as an anti-particle.

\bigskip\noindent
(iii) Finally, we look at $\Delta = 0$. Of course, in setting
$\Delta = 0$ we can no longer regard the equation as a gap equation
telling us how large $\Delta$ ought to be. What we can do, however,
is first, verify that for general $m$, $\mu_e$ and
$\mu_{\nu}^{\prime}$ the channel is attractive, and second,
discover under what restrictions on $m, \mu_e$ and
$\mu_{\nu}^{\prime}$ we can obtain a pairing instability. In this
case,

\begin{equation}
{\cal G}_0 (p_0, p) = {p_0 - \mu_e + p \over [(p_0 - \mu_e)^2 - p^2
- m^2] [p_0 + \mu_{\nu}^{\prime} + p]}
\end{equation}

\bigskip\noindent
and doing the $p_0$ integral we find, with $\omega = \sqrt{p^2 +
m^2}$,

\begin{eqnarray}
1 = {\kappa^2 \over \pi^2} \int_{-
\Lambda}^{\Lambda} ~dp ~p^2 \{{\theta(\omega^2 - \mu_e^2) \over \omega}
[{(\omega - p) \theta (-p - \mu_{\nu}^{\prime}
) \over \omega - p - \mu_e - \mu_{\nu}^{\prime}} + {(\omega + p)
\theta(p + \mu_{\nu}^{\prime}) \over \omega + p + \mu_e +
\mu_{\nu}^{\prime}}]  \nonumber \\
+ {2 (\mu_e + \mu_{\nu}^{\prime}) \over (\omega + p + \mu_e +
\mu_{\nu}^{\prime}) (\omega - p - \mu_e - \mu_{\nu}^{\prime})} [\theta(-
\mu_e - \omega) \theta (- p - \mu_{\nu}^{\prime}) - \theta (\mu_e
 + \omega) \theta (p + \mu_{\nu}^{\prime})]\} ~.
 \end{eqnarray}

\bigskip\noindent
Careful consideration of the $\theta$-functions reveals that the
integrand is always positive, thereby assuring us that the channel
is always attractive. The only conditions under which a denominator
can vanish, however, are

\begin{equation}
p^2 = \mu_{\nu}^{\prime 2}= \mu_e^2 - m^2 ~.
\end{equation}

\bigskip\noindent
This pair of conditions tells us that the fermi momenta of the two
members of the pair must coincide in order for an instability to
develop.

\vfill\eject
\noindent
REFERENCES

\noindent
1. J. Kepler, "Mysterium Cosmographicum", Tuebingen (1596).

\noindent
2. J. Balmer, Verh. Naturf. Ges. Basel {\bf 7}, 548 (1885).

\noindent
3. A.S. Eddington, "Relativity Theory of Protons and Electrons",
Cambridge University Press, Cambridge, UK (1936).

\noindent
4. P.A.M. Dirac, Proc. Roy. Soc. London {\bf A165}, 199(1938);
ibid. {\bf A365}, 19 (1979).

\noindent
5. A. Wyler, Acad. Sci. Paris, Comptes Rendus {\bf 269A}, 743
(1969).

\noindent
6. R. Roskies, Physics Today, Nov. 1971, p. 9; A. Peres, ibid.

\noindent
7. D.G. Caldi and A. Chodos, hep-ph/9903416.

\noindent
8. A. Chodos and E. Poppitz, manuscript in preparation.

\noindent
9. For reviews, see J.W.F. Valle, hep-ph/9907222; E.
Torrente-Lujan, hep-ph/9902339; B. Kayser, hep-ph/9810513.

\noindent
10. P.B. Pal and K. Kar, hep-ph/9809410. See also W.H. Kinney and
A. Riotto, hep-ph/9903459 and J. Lesgourgues and S. Pastor,
hep-ph/9904411.

\noindent
11. S. Perlmutter, et. al., astro-ph/9812473 and astro-ph/9812133;
B. Schmidt, et al., Astrophys. J. {\bf 507}, 46 (1998); A.G. Riess,
et al., astro-ph/9805201; Krauss, L.M. and Turner, M.S., Gen. Rel.
Grav. {\bf 27}, 1137 (1995); Ostriker, J.P. and Steinhardt, P.,
Nature {\bf 377}, 600 (1995).

\noindent
12. R.R. Caldwell, R. Dave and P.J. Steinhardt, Phys. Rev. Lett.
{\bf 80}, 1582 (1998).

\noindent
13. P.D. Mannheim, astro-ph/9901219.

\noindent
14. N. Arkani-Hamed, S. Dimopoulos and G. Dvali, Phys. Lett. {\bf
B429}, 263 (1998).

\noindent
15. A. Chodos, F. Cooper and H. Minakata, hep-ph/9905521 and A.
Chodos, H. Minakata and F. Cooper, Phys. Lett. {\bf B449}, 260
(1999).

\noindent
16. D.J. Gross and A. Neveu, Phys. Rev. {\bf D10}, 3235 (1974).

\noindent
17. H. Ishihara, K. Tomita and H. Nariai, Prog. Theor. Phys. {\bf
71}, 859 (1984).

\end{document}